\documentstyle[12pt]{article}

\title{\bf Structure Functions and Particle Production
in the Cumulative Region: two different exponentials
\thanks{Supported by the Russian Foundation for Fundamental
Research, Grant No. 94-02-06024-a}
}
\author{M.Braun and V.Vechernin \\
{\it Department of high-energy physics,}\\
{\it University of S. Petersburg, 198904 S. Petersburg, Russia}}
\date{}

\def\bc{\begin{center}}
\def\ec{\end{center}}
\def\beq{\begin{equation}}
\def\eeq{\end{equation}}

\def\v#1#2#3#4{
\put(#1,#2){\circle*{2}}
\put(#1,#3){\circle*{2}}
\put(#1,#2){\dashbox{1.5}(0,#4)[t]{}}
}

\begin{document}
\maketitle
\medskip

\begin{abstract}
In the framework of the recently proposed QCD based parton model for the
cumulative phenomena in the interactions with nuclei two mechanisms for
particle production, direct and spectator ones, are analysed. It is
shown that due to final state interactions the leading terms of the direct
mechanism contribution are cancelled and the spectator mechanism is the
dominant one. It leads to a smaller slope of the cumulative particle
production rates compared to the slope of the nuclear structure function
in the cumulative region $x>1$, in agreement with the recent
experimental data.
\end{abstract}

\section{Introduction}

We have recently proposed a QCD based parton model for the cumulative
phenomena in the interactions with nuclei [1]. In the model quark
diagrams are summed in the vicinity of all intermediate thresholds
at which some quarks of the nucleus ("donors") transfer all their
longitudinal momenta to the distinguished active quark and become soft.
Particle production proceeds in the model via two different mechanisms,
the direct and spectator ones, schematically shown in Fig.~1 $a,b$.,
respectively. The simpler direct mechanism leads to the same $x$
dependence of the production rate $I_{d}$ as for the structure function
$F_{2}(x)$ in the region $x>1$. It is roughly an exponential in $x$
\beq
I_{d}(x)\sim F_{2}(x)\sim\exp(-b_{0}x)
\eeq
where the slope $b_{0}$ is determined by the QCD coupling constant and
quark mass. The spectator mechanism besides involves interactions
between partons of the projectile and target nucleus. It was shown in
[1] that each donor quark has to interact with the projectile. As a
result the spectator contribution $I_{s}$ also behaves in $x$ as an
exponential but with a different slope
\beq
I_{s}(x)\sim\exp (-b_{s}x)
\eeq
the slope $b_{s}$ depending not only on the QCD coupling constant and
quark mass but also on the partonic amplitude. With a particular
parametrization  for the latter chosen in [1] the slope $b_{s}$ was
found to be somewhat larger than $b_{0}$, so that the spectator
contribution resulted smaller than the direct one. Then the total
particle production rate behaves in $x$ exactly as the structure
function. At the time when the paper [1] was written there existed no
reliable data on the cumulative structure functions except the old data
on the deuterium at low energies and not really in the deep inelastic
region [2]. These deuterium data had the slope in $x$ much steeper
than particle production data from [3,4]. Assuming the direct
mechanism for particle production this resulted in very different values
for the effective coupling constant found in [1] for the two
cumulative phenomena. In view of a preliminary character of the
deuterium data we preferred not to pursue this point any further in [1].

However recently a new set of data on the structure function of
$^{12}{\rm C}$ in the region $0.8\leq x\leq 1.3$ was published [5],
which essentially agrees with the $x$ dependence
observed on the deuterium in [2].
Thus it seems to be confirmed that the structure function and
particle production have  different $x$ dependence in the cumulative
region. The slope $b_{0}$ as observed in [2] and [5] is roughly
16, whereas the slope $b_{p}$ of the particle production rate is of the
order 6$\div$8 [3,4].

This circumstance gives us  a motivation to reconsider our study of the
cumulative phenomena in the light of the experimental evidence for two
different exponentials for the structure function and particle
production. In the framework of our model it means that it is the
spectator mechanism that gives the bulk of particle production.

In this note we first demonstrate that  final state interactions, not
taken into account in [1], cancel the leading terms in the direct
contribution, so that it becomes much less than the spectator one. Then
particle production in the cumulative region indeed goes predominantly
via the spectator mechanism. To be able to explain the experimental slope
we study a wider class of parametrizations for the partonic amplitude
than in [1]. The spectator slope $b_{s}$ results very sensitive to the
magnitude of the hadronic diffractive cross-sections. The
parametrization
used in [1], which lead to a rather large value for $b_{s}\sim 15$,
corresponded to a zero diffractive cross-section. Raising the ratio of
the diffractive to elastic cross-section up to 1.5 results in lowering
$b_{s}$ down to 7$\div$9, not very far from the experimentally
observed value
6$\div$8. However it does not seem possible to push $b_{s}$ still
further down without entering into a serious conflict with the data on
hadronic cross-sections.

Note that a more phenomenological attempt to explain smaller slopes for
the cumulative particle production is made in [6], where the existence
of multiquark clusters in the nucleus  and their properties are
postulated and the quark-gluon string model formalism of [7] is used
to calculate production rates.

The paper is organized as follows. In
Sec.~2 we show how the leading terms in the direct contribution
are cancelled by final state
interactions. In Sec.~3 the partonic amplitude is related to the
hadronic one and the data on hadronic cross-sections are used to
study its possible parametrizations. In Sec.~4 we calculate the particle
production rate by the spectator mechanism and compare its $x$
dependence to that of the structure function and the experimentally
observed one.

\section{Cancellation of the direct contribution}

In the upper blob of Fig.~1 $a$ the inclusive cross-section off the
active quark enters in the triple Regge kinematical region, since the
scaling variable $x$ of the produced particle should be as close as
possible to that of the active quark. Therefore we can redraw the
diagram of Fig.~1 $a$ in the form shown in Fig.~2 $a$. However we
have also to take into account final state interactions of the produced
particle with donor quarks which raise its longitudinal momentum. An
example of these is shown in Fig.~2 $b$. The diagrams of Fig.~2 $a$ and
$b$ are not the only ones which give contribution. In fact taking a
diagram with a certain pattern of interactions with donor quarks
one should sum over all places where the two upper reggeons
from the triple Regge interaction are attached to the active quark. We
are going to demonstrate that this sum vanishes in the limit when the
donor quarks loose all their longitudinal momentum.

Let the number of donors be $n$. We denote the momenta of the left
active quark $\kappa_{0},...\kappa_{l}$ before the interaction with the
reggeon and $\kappa'_{l},\kappa'_{l+1}...\kappa'_{n}\equiv\kappa$
after this interaction.
Analogous quantities on the right side hereafter will be
labelled with tildas (see Fig.~2 $a,b$). We use the light-cone variables
$k=(k_{\pm},k_{\bot})$. The donor quarks lie on the mass shell between
the interactions. So there minus momenta ("energies") take on physical
values
\beq
k_{i-}=(m^{2}-k_{i\bot}^{2})/2k_{i+}\equiv\mu_{i}
\eeq
From energy conservation we find for the energies of the active quark
\beq
\kappa_{i-}=Np_{1-}-\sum_{n+1}^{N-1}\mu_{j}
-\sum_{1}^{i}\mu_{j}-\sum_{i+1}^{n}k_{j-}^{(0)},\ \ i=1,2,...l
\eeq
and
\beq
\kappa'_{i-}=\kappa_{-}+\sum_{i+1}^{n}(\mu_{j}-k_{j-}^{(0)})
\ \ i=l,l+1,...n-1
\eeq
where $N$ is the total number of partons in the nucleus and
$k_{i}^{(0)}$ are the momenta of the left donors before the interaction.
Integrations over $k_{i-}^{(0)}$ can then be done easily, since all
singularities coming from the active quark propagators lie on the
opposite side of the real axis as compared to the donor propagators.
As a result $k_{i-}^{(0)}$ are substituted by $\mu_{i}^{(0)}$ in (4) and
(5). In the limit when the donor quarks loose all there longitudinal
momenta $\mu_{i}$ become large and from (4) and (5) one obtains for the
$i$th active quark propagator
\beq
1/\sum_{1}^{i}\mu_{j}, \ \ i=1,2,...l
\eeq
and
\beq
-1/\sum_{i+1}^{n}\mu_{j}, \ \ i=l,l+1,...n-1
\eeq

Now we pass to the triple Regge interaction. The momentum of the upper
left reggeon $q$ is
\beq
q=\kappa'_{l}-\kappa_{l}
\eeq
So $q_{-}=\sum_{1}^{n}\mu_{i}$. Calculating $q_{+}$ and $q_{\bot}$ we
find
\beq
q^{2}=-2p_{1+}(\Delta-\sum_{1}^{n}x_{i})\sum_{1}^{n}\mu_{i}+
(\kappa+\sum_{1}^{N-1}k_{i})_{\bot}^{2}
\eeq
Here $x_{i}$ are the scaling momenta of the donors after the
interactions, $x_{i}\rightarrow 0$; $\Delta=n+1-x$ where $x$
is the scaling variable of the produced particle. All scaling variables
are defined here relative to the average longitudinal momentum of the
initial quarks of the nucleus. They are three times larger than the
standard ones defined relative to the nucleon momentum. At the threshold
$\Delta\rightarrow 0$. The important thing about (9) is that $q^{2}$ does
not depend on $l$, that is, on the place at which the reggeon is
attached.

However there still remains some dependence on $l$ in the energy on
which the upper reggeons depend. In fact the factor corresponding to the
triple Regge interaction is given by
\beq
g(t)(s_{l}\tilde{s}_{\tilde{l}})^{\alpha(t)}(M^{2})^{\alpha (0)-2\alpha(t)},
\ \ t=q^{2}
\eeq
Here $q^{2}$ is given by Eq.~(9), $g(t)$ is the three-reggeon vertex,
$M^{2}=(\Delta-\sum_{1}^{n}x_{i})s$ where $s$ is the standard energy
variable; $s_{l}=\xi_{l}s$ where $\xi_{l}$ is the scaling variable of
the left active quark at the moment of its interaction with the reggeon
and $\tilde{s}_{\tilde{l}}$ is a similar quantity on the right. All the three
reggeons are taken to be pomerons with the trajectory $\alpha(t)$. We
have also
to take into account that the $l$th propagator of the active quark is
splitted into two by the interaction with the reggeon, which results
in an
extra factor $1/\xi_{l}$. Introducing it into (10) we finally obtain for
the triple Regge interaction factor
\beq
g(t)s^{\alpha (0)}(\xi_{l}\tilde{\xi}_{\tilde{l}})^{\alpha(t)-1}
(\Delta-\sum_{1}^{n}x_{i})^{\alpha(0)-2\alpha(t)}
\eeq

If we assume  weak (logarithmic) dependence of the cross-section on
energy then we have to take for the effective pomeron intercept
$\alpha(0)=1$. Then the dependence of the expression (11) on $l$ and
$\tilde{l}$ is also weak (it enters only through the term
$\xi_{l}^{\alpha't}$ with small $\alpha'$ and $t$). Neglecting this
dependence we find that the triple Regge interaction factor does not
depend on the place where it is attached and is common to all the
diagrams of the type shown in Fig.~2.

Then we have only to sum various diagrams forgetting about the common
triple Regge factor. Diagrams with different $l$ (or/and
$\tilde{l}$) differ only by their active quark propagators. For a given
$l$ the factor coming from the left active quark propagators is found
from (7) and (8) to be
\beq
\prod_{i=1}^{l}(1/\sum_{1}^{i}\mu_{j})
\prod_{i=l}^{n-1}(-1/\sum_{i+1}^{n}\mu_{j})
\eeq
Upon symmetrizing in the soft donors it becomes
\beq
(l!(n-l)!)^{-1}(-1)^{n-l}\prod_{1}^{n}1/\mu_{i}
\eeq
The sum of (13) over all $l=0,1,...n$ evidently gives zero. Thus in the
described approximation when the slow change of the triple reggeon
interaction with the energy of the active quark is neglected all the
diagrams of the type shown in Fig.~2 cancel in the sum.

\section{Parton interaction and hadronic cross-sections}

With the bulk of the direct contribution cancelled, particle production
in the cumulative region goes predominantly via the spectator mechanism
of Fig.~1 $b$. It involves interactions between partons of the
projectile and donors from the nucleus, characterized by the partonic
amplitude $a$. We normalize it according to
\beq
2 {\rm Im}\,a(0)=\sigma
\eeq
where $a(0)$ is the forward amplitude and $\sigma$ is the partonic
cross-section. As shown in [1] each donor parton has to take part in
the interaction with the projectile. With $n$ donors, this gives a
factor $a^{n}$ which is responsible for the difference between the
slopes of the structure function and particle production. The
parametrization of $a$ and its magnitude thus aquire the decisive role
in describing the experimental slope.

Of course, $a$ is not the quantity to be directly measured
experimentally. It can however be related to the data through hadronic
($pp$ or $p\bar p$) interaction [8]. The elastic hadronic amplitude
can be represented through partonic interaction as shown in the diagram
of Fig.~3. Taking the c.m. system one finds that the longitudinal
components
$q_{i\pm}$ of the transferred momenta are small. Also the upper part of
the diagram belonging to the projectile does not depend on $q_{i+}$ and
the lower, target part does not depend on $q_{i-}$. Integrating over
$q_{i\pm}$ one then finds that for a diagram with $n$ interactions,
$M$ partons in the projectile and $N$ ones in the target the amplitude
is given by
\beq
iA_{n}^{(M,N)}(q_{\bot})=n!C_{M}^{n}C_{N}^{n}\int\prod_{1}^{n}
(\frac{d^{2}q_{i\bot}}{(2\pi )^{2}}ia(q_{i\bot}))(2\pi )^{2}
\delta^{2}(q-\sum_{1}^{n}q_{i})F_{n}^{(M)}(q_{i\bot})
 F_{n}^{(N)}(q_{i\bot})
\eeq
Here $F_{n}^{(M,N)}(q_{i\bot})$ are the $n$-fold transverse form-factors
for the projectile ($M$) and target ($N$). Their Fourier transforms give
parton distributions in the transverse space $F_{n}^{(M,N)}(b_{i})$.
The amplitude (15) thus can be presented as an integral in the transverse
space
\beq
iA_{n}^{(M,N)}(q_{\bot})=n!C_{M}^{n}C_{N}^{n}\int
d^{2}B\exp (iqB)
\prod_{1}^{n}
(d^{2}b_{i}d^{2}b'_{i}a(B-b_{i}+b'_{i}))
F_{n}^{(M)}(b_{i}) F_{n}^{(N)}(b'_{i})
\eeq
This is only a contribution from given numbers of partons in the
projectile and target. Summing over $M$ and $N$ we obtain for the
elastic amplitude with $n$ partonic interactions
\beq
iA_{n}(q_{\bot})=(1/n!)\int
d^{2}B\exp (iqB)
\prod_{1}^{n}
(d^{2}b_{i}d^{2}b'_{i}a(B-b_{i}+b'_{i}))
F_{n}(b_{i}) F_{n}(b'_{i})
\eeq
where for the projectile
\beq
F_{n}(b_{i})=\sum_{M\geq n}(M!/(M-n)!) F_{n}^{(M)}(b_{i})
\eeq
and similarly for the target.

To move further one has to make some assumptions about the multiparton
distributions $F_{n}(b_{i})$. The simplest one is to assume that partons
are completely independent: their number is distributed according to
Poisson's law and multiparton distributions factorize. One then obtains
\beq
F_{n}(b_{1},...b_{n})=\nu^{n}\prod_{1}^{n}\rho(b_{i})
\eeq
where $\nu$ is the mean number of partons in the projectile or target
and the one parton distribution $\rho$ is normalized to unity
\beq
\int d^{2}b\rho (b)=1
\eeq

Putting (19) into (17) and summing over $n$ one arrives at an eikonal
amplitude
\beq
iA(q_{\bot})=\int
d^{2}B\exp (iqB) (\exp (ip(B))-1)
\eeq
where the eikonal factor is
\beq
p(B)=\nu_{p}\nu_{t}\int d^{2}bd^{2}b'\rho_{p}(b)\rho_{t}(b')
a(B-b+b')
\eeq
and the subscripts $p$ and $t$ refer to the projectile and target,
respectively. This form of the amplitude coincides with the one found in
the multipomeron exchange model with factorizable vertices [9].
Eq.~(22) then gives the contribution of a single pomeron exchange.

One commonly assumes that both $\rho$ and $a$ have a Gaussian dependence
on the impact parameter:
\beq
\rho_{p,t}(r)=(1/\pi r_{p,t}^{2})\exp(-r^{2}/r_{p,t}^{2}), \ \
a(r)=(i\sigma/2\pi r_{0}^{2})\exp(-r^{2}/r_{0}^{2})
\eeq
where we have taken $a$ pure imaginary, for simplicity. Then one easily
finds simple expressions for
the eikonal factor, the hadronic total, elastic and inelastic
cross-sections and also for the slope of the elastic cross-section
$B^{el}$:
\[p(B)=ix\exp (-B^2/R^2)\]
\[\sigma^{tot}=2\pi R^{2}\phi_{1}(x)\]
\[\sigma^{el}=\pi R^{2}(2\phi_{1}(x) -\phi_{1}(2x))\]
\[\sigma^{in}=\pi R^{2}\phi_{1}(2x)\]
\beq
B^{el}=(R^2/2)\phi_{2}(x)/\phi_{1}(x)
\eeq
where $R^{2}=r_{p}^{2}+r_{t}^{2}+r_{0}^{2}$ is the total interaction
radius squared, $x=\nu_{p}\nu_{t}\sigma/(2\pi R^{2})$ and the functions
$\phi_{1,2}(x)$ are defined by
\beq
\phi_{n}(x)=\sum_{k=1}^{\infty} (-1)^{k-1}x^{k}/(k!k^{n})
\eeq

Taking the ratio of any pair of the quantities in (24) and comparing it to
its experimental value one can determine $x$. Then one of Eqs.~(24)
can be used to find $R^{2}$. With $x$ and $R^{2}$ determined, the
cross-section $\nu_{p}\nu_{t}\sigma$ can be found. Values of $r_{p}$ and
$r_{t}$ are more or less known from electromagnetic properties of the
projectile and target, thus $r_{0}$ can also be found. So, in the end,
the only parameters left are the average numbers of partons in the
projectile and target $\nu_{p}$ and $\nu_{t}$. At low energies the
valence quark approximation seems to be good enough, which implies
$\nu_{p}=\nu_{t}=3$ for $pp$ or $p\bar p$ interactions.

This standard procedure was used in [1] to determine the parameters
entering the spectator mechanism. As mentioned, it lead to a relatively
small $a$, so that the slope of the spectator spectrum resulted even
steeper than that of the structure function. However, apart from this
unsatisfactory result for cumulative production, the discussed
parametrization has a more fundamental (and well-known) defect. As one
can deduce from the first three equations in (24) it gives no diffraction.
This property of the pure eikonal amplitude can be easily understood
in the framework of the Gribov approach to multiple scattering [10],
where it corresponds to retaining only the initial particle state in the
sum over all intermediate states. A possible remedy consists in changing
the eikonal amplitude (21) by the quasi-eikonal one with a diffraction
factor $\xi >1$ [11]:
\beq
iA(q_{\bot})=\int
d^{2}B\exp (iqB)\xi^{-1} (\exp (i\xi p(B))-1)
\eeq
With this factor Eqs.~(24) change as follows
\[\sigma^{tot}=\xi^{-1}2\pi R^{2}\phi_{1}(x)\]
\[\sigma^{el}=\xi^{-2}\pi R^{2}(2\phi_{1}(x) -\phi_{1}(2x))\]
\beq
\sigma^{in}=\xi^{-1}\pi R^{2}\phi_{1}(2x)
\eeq
and $B^{el}$ does not change at all. From (27) one finds the diffractive
cross-section
\beq
\sigma^{dif}=(\xi -1)\sigma^{el}
\eeq
so that the new parameter $\xi$ can be directly taken from
experiment.

In our partonic approach the quasi-eikonal parametrization (26) means that
instead of (19) we take
\beq
F_{n}(b_{1},...b_{n})=\xi^{(n-1)/2}\nu^{n}\prod_{1}^{n}\rho(b_{i})
\eeq
both for the target and projectile.

The quasi-eikonal parametrization leads to values of $a$ considerably
larger than without diffraction because of a stronger screening effect
introduced by diffractive states (see Table). For the maximum value of
$\xi\simeq 2.4$ compatible with the experimental data at $\sqrt{s}=
23.5\ GeV$ it leads to the value of the corresponding parton-nucleon
cross-section $\sigma_{1}$ nearly three times larger than with $\xi=1$
used in [1]. Note that the absolute value of $\sigma_{1}$ then results
quite large ($\sim 50\ mb$). It is interesting that values of
$\sigma_{1}$ of a similar order are also favoured by the study of the
behaviour of the nuclear structure functions in the opposite kinematical
region at small $x$ [12]. In relation to this we recall that the
parton-nucleon cross-section $\sigma_{1}$ is not a directly observable
quantity. It only enters hadronic cross-sections as a parameter. The
mentioned large value of $\sigma_{1}$ is consistent with all
experimental evidence on proton-proton interactions at $\sqrt{s}=
23.5\ GeV$. This large value transforms into the physical value of
the $pp$ cross-section as a result of a strong screening effect, that
is, because of a highly coherent manner in which the quarks interact.

\section{Numerical results and discussion}

With the partonic amplitude fixed in the preceding section, one can
calculate the spectator contribution to the cumulative production rate,
using  the coresponding equations from our paper [1] (Eqs.~(38), (39)
and (41) of [1]), suitably modified to the quasi-eikonal case. We
have also corrected a mistake eliminating $p!$ in the denominator of
Eq.~(39) of [1]. We have chosen
the maximum possible value for the diffractive parameter
$\xi=2.4$. Other parameters have been fixed as in [1].

The results of the calculations are shown in Figs.~4 and 5 for the
cumulative charged pion production on deuterium and  $^{181}$Ta,
respectively. For comparison the corresponding nuclear structure
function $F_{2}(x)$ at $x>1$, calculared in [1], is also shown, as
well as the available experimental data from [2-5].

One clearly observes that the spectaor mechanism, with a parametrization
of the partonic amplitude chosen to account for diffraction, leads to a
considerably smaller slope of the production spectra ($b_{s}\sim 7\div 9$,
see Introduction) compared to the slope of the structure function in the
region $x>1$ ($b_{0}\sim 16$), in a good agreement with experimental
data.

Thus our model correctly predicts two differents exponential in the
cumulative production and in the nuclear structure function at $x>1$.
The difference is due to additional multiple interactions between
projectile and target which enter the spectator mechanism for the
cumulative production.

\newpage
\vspace*{3 cm}
{\Large\bf Table. Cross-sections and elastic slopes for
various values of the diffractive factor $\xi$}
\vspace {1 cm}

\begin{tabular}{|r|r|r|r|}\hline
$\xi$ &$\sigma^{dif}$ ({\it mb})&$\sigma_{1}$ ({\it mb})&
$B^{el}$ ($ GeV^{-2}$)\\\hline
1.0  & 0.   & 16.5  &  11.1  \\\hline
1.5  & 3.4  & 20.2  &  10.9  \\\hline
2.0  & 6.8  & 31.2  &  10.6  \\\hline
2.2  & 8.2  & 44.3  &  10.6  \\\hline
2.4  & 9.5  & 48.3  &  11.4  \\\hline
\end{tabular}
\vspace{1.5 cm}

{\Large\bf Table captions}
\vspace{0.5 cm}

The first column gives  values of the diffractive parameter $\xi$.
In the second column the corresponding  $pp$ diffractive
cross-section at $\sqrt{s}=23.5\ GeV$ is shown. In the third column the
values of  the parton-projectile
(proton) cross section are shown, as calculated from Eqs.~(24) (from the
value of  $x$ with $\nu=3$).
The last column shows the resulting  elastic slope,
its experimental value at $\sqrt{s}=23.5\ GeV$ being $B^{(el)}=
11.8\pm 0.30\ GeV^{-2}$ [13].

\newpage

{\large {\bf Figure captions}}

\begin{description}

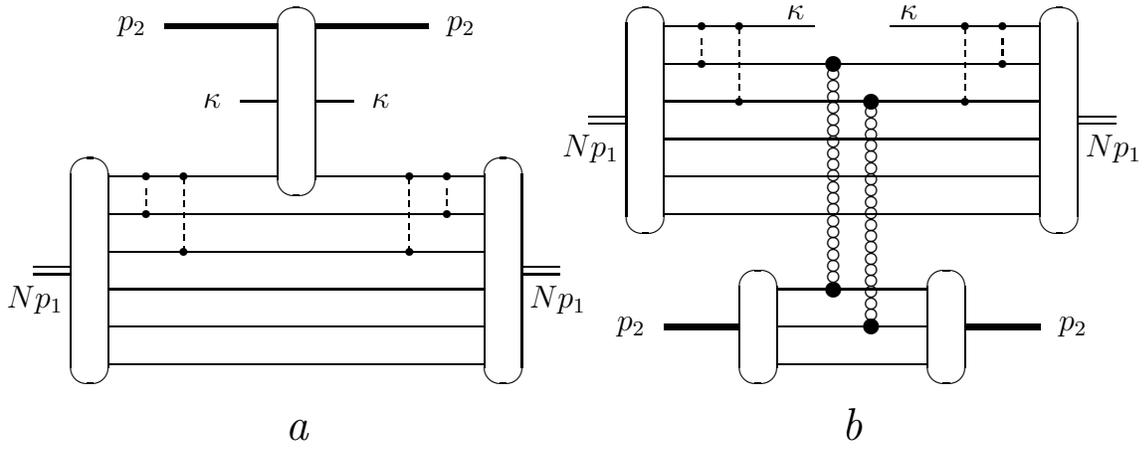
\item[Fig.~1]
Two mechanisms of the cumulative particle production
in the parton model,
the direct ($a$) and spectator ($b$) ones.
Dashed and chain lines show gluon and pomeron
exchanges, respectively.

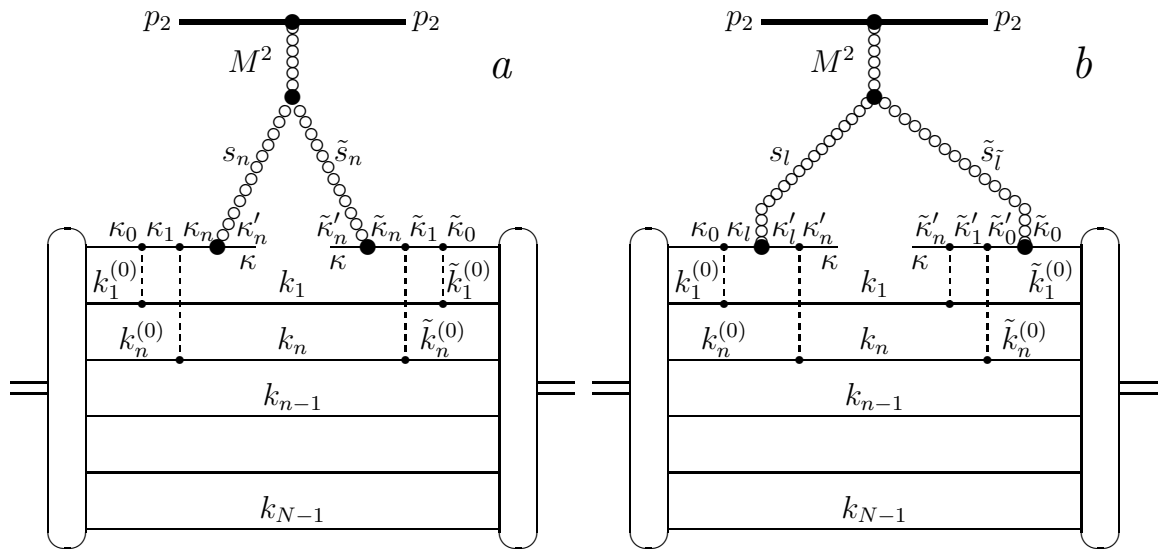
\item[Fig.~2]
The direct mechanism of the cumulative particle production
in the triple Regge approach without ($a$) and with ($b$)
final state interactions of the produced
particle with donor quarks.
Shown for the cases ($n=l=\tilde{l}=2$) and
($n=2, l=1, \tilde{l}=0$), respectively.
Notations as in Fig.~1.

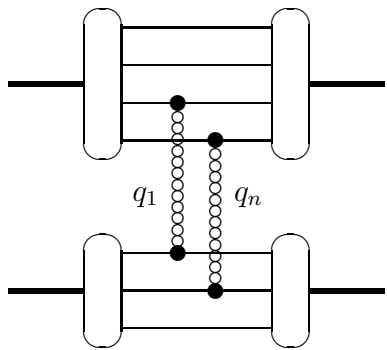
\item[Fig.~3]
The elastic hadronic amplitude
in the parton model.
Notations as in Fig.~1.

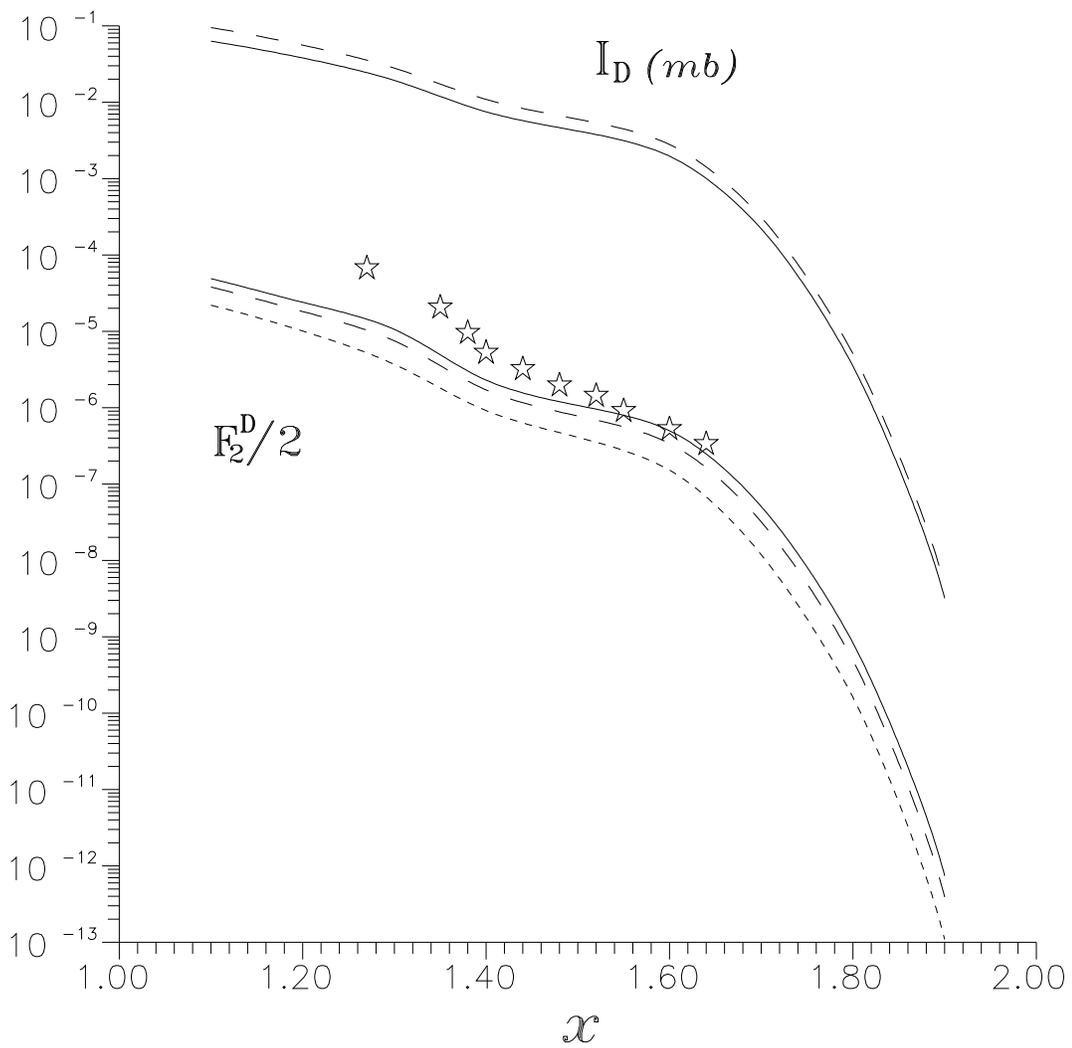
\item[Fig.~4]
$I_D=\frac{xd\sigma}{2dx}(mb)$ is
the calculated inclusive cross-section (per nucleon)
for cumulative charged pion production on deuteron
at  $\sqrt{s}=23.5\ GeV$ (solid curve) and
$1800\ GeV$ (dashed one).
$F^D_2/2$ is
the calculated structure function of the
deuteron (per nucleon) for $x>1$
at  $Q^2=6\ GeV^2$ (solid curve),
$20\ GeV^2$ (dashed curve) and
$500\ GeV^2$ (short dash curve).
$\star$ - the experimental data [2] on the
deuteron structure function
at  $0.8 \leq Q^2 \leq 6\ GeV^2$.
The experimental errors are much less than the star symbols
and not discernable on our scale.

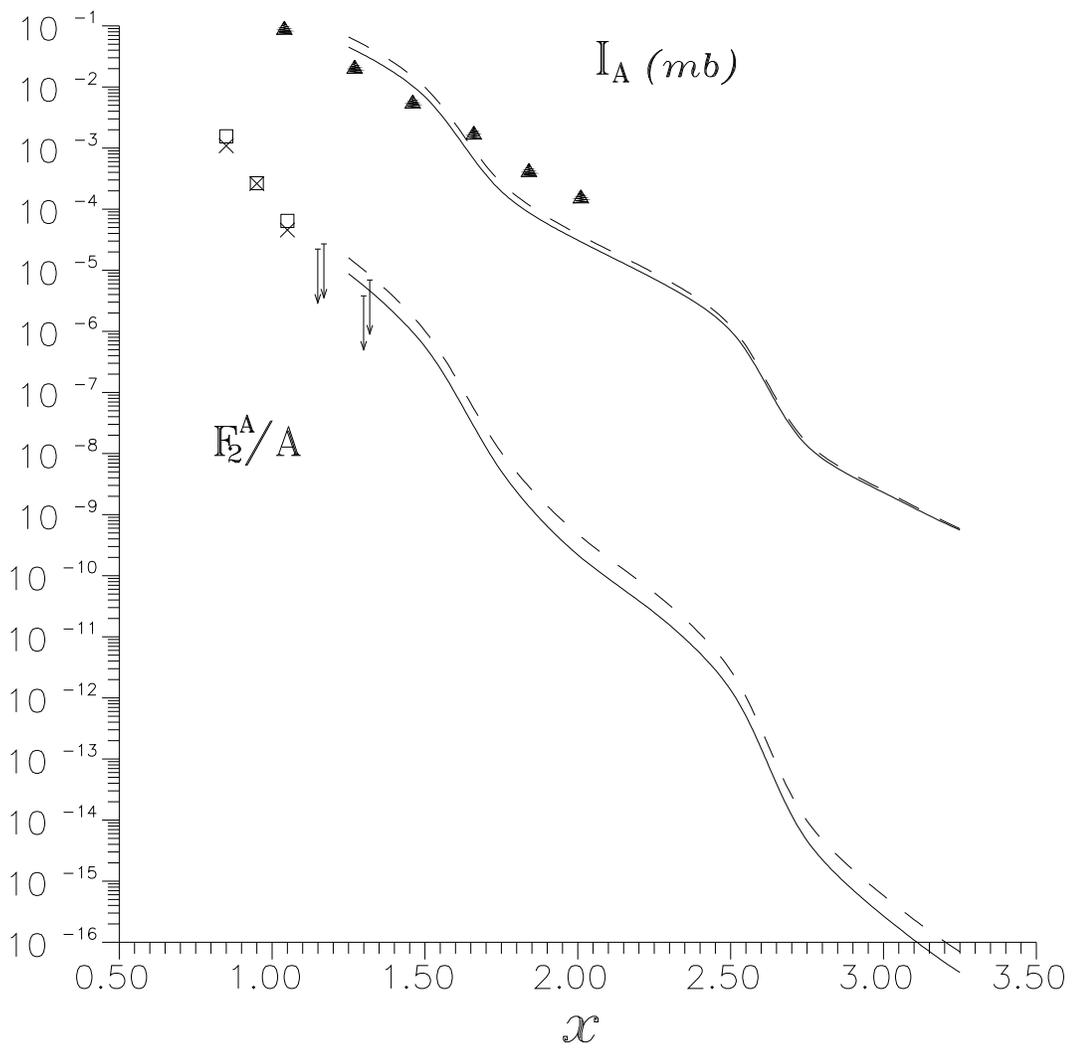
\item[Fig.~5]
$I_A(mb)$ is the same as in Fig.~4
but for the production on $^{181}{\rm Ta}$.
$\Delta$ - the experimental data [4] on the
cumulative charged pion production on
$^{181}{\rm Ta}$
by $400\ GeV$ incident proton beam.
$F^A_2/2$ is
the calculated nuclear structure function for the
$^{181}{\rm Ta}$ at $Q^2=50\ GeV^2$ (dashed curve) and
$^{12}{\rm C}$ at $Q^2=100\ GeV^2$ (solid curve).
$\Box$ and $\times$ - the experimental data [5] on the
$^{12}{\rm C}$ structure function
at $Q^2=61\ GeV^2$ and  $150\ GeV^2$, respectively.

\end{description}

\newpage
\vspace*{5cm}
\unitlength=0.5mm
\linethickness{0.4pt}
\noindent
\begin{picture}(70,140)(-80,-20)

\put(0,-10){\makebox(0,0)[cb]{\Large{\it a}}}

\def\fs1{
\v{30}{40}{60}{20}
\v{40}{50}{60}{10}
\v{-30}{40}{60}{20}
\v{-40}{50}{60}{10}

\put(-50,10){\line(1,0){100}}
\put(-50,20){\line(1,0){100}}
\put(-50,30){\line(1,0){100}}
\put(-50,40){\line(1,0){100}}
\put(-50,50){\line(1,0){100}}

\put(55,35){\oval(10,60)[]}
\put(60,34){\line(1,0){10}}
\put(60,36){\line(1,0){10}}
\put(-55,35){\oval(10,60)[]}
\put(-70,34){\line(1,0){10}}
\put(-70,36){\line(1,0){10}}

\put(62,31){\makebox(0,0)[lt]{$Np_1$}}
\put(-62,31){\makebox(0,0)[rt]{$Np_1$}}
}

\fs1

\put(0,80){\oval(10,50)[]}
\linethickness{2.pt}
\put(5,100){\line(1,0){30}}
\put(-5,100){\line(-1,0){30}}
\linethickness{0.4pt}
\put(40,100){\makebox(0,0)[lc]{$p_2$}}
\put(-40,100){\makebox(0,0)[rc]{$p_2$}}

\put(5,80){\line(1,0){10}}
\put(-5,80){\line(-1,0){10}}
\put(20,80){\makebox(0,0)[lc]{$\kappa$}}
\put(-20,80){\makebox(0,0)[rc]{$\kappa$}}

\put(5,60){\line(1,0){45}}
\put(-5,60){\line(-1,0){45}}

\put(145,40){
\begin{picture}(70,150)

\put(0,-50){\makebox(0,0)[cb]{\Large{\it b}}}
\fs1

\put(-50,60){\line(1,0){40}}
\put(10,60){\line(1,0){40}}

\multiput(-5,-10)(0,3){20}{\makebox(0,0)[cc]{$\circ$}}
\multiput(5,-20)(0,3){20}{\makebox(0,0)[cc]{$\circ$}}
\put(-5,-10){\circle*{4}}
\put(-5,50){\circle*{4}}
\put(5,-20){\circle*{4}}
\put(5,40){\circle*{4}}

\put(25,-20){\oval(10,30)[]}
\put(-25,-20){\oval(10,30)[]}
\linethickness{2.pt}
\put(30,-20){\line(1,0){20}}
\put(-30,-20){\line(-1,0){20}}
\linethickness{0.4pt}
\put(55,-20){\makebox(0,0)[lc]{$p_2$}}
\put(-55,-20){\makebox(0,0)[rc]{$p_2$}}

\put(-20,-10){\line(1,0){40}}
\put(-20,-20){\line(1,0){40}}
\put(-20,-30){\line(1,0){40}}

\put(15,62){\makebox(0,0)[cb]{$\kappa$}}
\put(-15,62){\makebox(0,0)[cb]{$\kappa$}}

\end{picture}
}
\end{picture}

\vfill

\bc
Fig.~1
\ec

\newpage
\unitlength=0.5mm
\linethickness{0.4pt}
\noindent
\begin{picture}(70,150)(-75,-25)

\v{30}{30}{60}{30}
\v{40}{45}{60}{15}
\v{-30}{30}{60}{30}
\v{-40}{45}{60}{15}

\def\fs2{
\multiput(0,100)(0,3){7}{\makebox(0,0)[cc]{$\circ$}}

\put(-55,60){\line(1,0){45}}
\put(10,60){\line(1,0){45}}

\put(-55,-15){\line(1,0){110}}
\put(-55,0){\line(1,0){110}}
\put(-55,15){\line(1,0){110}}
\put(-55,30){\line(1,0){110}}
\put(-55,45){\line(1,0){110}}

\put(60,22.50){\oval(10,85)[]}
\put(-60,22.50){\oval(10,85)[]}
\put(65,21){\line(1,0){10}}
\put(65,24){\line(1,0){10}}
\put(-75,21){\line(1,0){10}}
\put(-75,24){\line(1,0){10}}

\put(0,100){\circle*{4}}
\put(0,120){\circle*{4}}

\put(-47,48){\makebox(0,0)[cb]{$k_1^{(0)}$}}
\put(-40,33){\makebox(0,0)[cb]{$k_n^{(0)}$}}
\put(47,48){\makebox(0,0)[cb]{$\tilde{k}_1^{(0)}$}}
\put(40,33){\makebox(0,0)[cb]{$\tilde{k}_n^{(0)}$}}

\put(0,48){\makebox(0,0)[cb]{$k_1$}}
\put(0,33){\makebox(0,0)[cb]{$k_n$}}
\put(0,18){\makebox(0,0)[cb]{$k_{n-1}$}}
\put(0,-12){\makebox(0,0)[cb]{$k_{N-1}$}}

\put(-32,120){\makebox(0,0)[rc]{$p_2$}}
\put(32,120){\makebox(0,0)[lc]{$p_2$}}
\put(-5,110){\makebox(0,0)[rc]{$M^2$}}

\put(-12,58){\makebox(0,0)[ct]{$\kappa$}}
\put(12,58){\makebox(0,0)[ct]{$\kappa$}}

\linethickness{2.pt}
\put(-30,120){\line(1,0){60}}
\linethickness{0.4pt}
}

\fs2

\multiput(-20,60)(1.5,3){13}{\makebox(0,0)[cc]{$\circ$}}
\multiput(20,60)(-1.5,3){13}{\makebox(0,0)[cc]{$\circ$}}

\put(55,105){\makebox(0,0)[cb]{\Large{\it a}}}

\put(-20,60){\circle*{4}}
\put(20,60){\circle*{4}}

\put(-45,63){\makebox(0,0)[cb]{$\kappa_0$}}
\put(-35,63){\makebox(0,0)[cb]{$\kappa_1$}}
\put(-25,63){\makebox(0,0)[cb]{$\kappa_n$}}
\put(45,63){\makebox(0,0)[cb]{$\tilde{\kappa}_0$}}
\put(35,63){\makebox(0,0)[cb]{$\tilde{\kappa}_1$}}
\put(25,63){\makebox(0,0)[cb]{$\tilde{\kappa}_n$}}

\put(-15,63){\makebox(0,0)[lb]{$\kappa'_n$}}
\put(15,63){\makebox(0,0)[rb]{$\tilde{\kappa}'_n$}}

\put(-11,83){\makebox(0,0)[rb]{$s_n$}}
\put(11,83){\makebox(0,0)[lb]{$\tilde{s}_n$}}

\put(152,0){
\begin{picture}(70,150)

\v{20}{45}{60}{15}
\v{30}{30}{60}{30}
\v{-20}{30}{60}{30}
\v{-40}{45}{60}{15}

\fs2

\multiput(-30,70)(2,2){15}{\makebox(0,0)[cc]{$\circ$}}
\multiput(40,70)(-2.66,2){15}{\makebox(0,0)[cc]{$\circ$}}
\multiput(-30,61)(0,3){3}{\makebox(0,0)[cc]{$\circ$}}
\multiput(40,61)(0,3){3}{\makebox(0,0)[cc]{$\circ$}}

\put(55,105){\makebox(0,0)[cb]{\Large{\it b}}}

\put(-30,60){\circle*{4}}
\put(40,60){\circle*{4}}

\put(-45,63){\makebox(0,0)[cb]{$\kappa_0$}}
\put(-36,63){\makebox(0,0)[cb]{$\kappa_l$}}
\put(-24,63){\makebox(0,0)[cb]{$\kappa'_l$}}
\put(46,63){\makebox(0,0)[cb]{$\tilde{\kappa}_0$}}
\put(34,63){\makebox(0,0)[cb]{$\tilde{\kappa}'_0$}}
\put(25,63){\makebox(0,0)[cb]{$\tilde{\kappa}'_1$}}

\put(-15,63){\makebox(0,0)[cb]{$\kappa'_n$}}
\put(15,63){\makebox(0,0)[cb]{$\tilde{\kappa}'_n$}}

\put(-22,83){\makebox(0,0)[rb]{$s_l$}}
\put(28,83){\makebox(0,0)[lb]{$\tilde{s}_{\tilde{l}}$}}

\end{picture}}

\end{picture}

\vspace{1cm}

\bc
Fig.~2
\ec

\vfill

\unitlength=0.5mm
\linethickness{0.4pt}
\bc
\begin{picture}(100,90)(-50,-35)

\put(25,35){\oval(10,40)[]}
\put(-25,35){\oval(10,40)[]}
\linethickness{2.pt}
\put(30,35){\line(1,0){20}}
\put(-30,35){\line(-1,0){20}}
\linethickness{0.4pt}
\put(-20,20){\line(1,0){40}}
\put(-20,30){\line(1,0){40}}
\put(-20,40){\line(1,0){40}}
\put(-20,50){\line(1,0){40}}

\multiput(-5,-10)(0,3){13}{\makebox(0,0)[cc]{$\circ$}}
\multiput(5,-20)(0,3){13}{\makebox(0,0)[cc]{$\circ$}}
\put(-5,-10){\circle*{4}}
\put(-5,30){\circle*{4}}
\put(5,-20){\circle*{4}}
\put(5,20){\circle*{4}}
\put(10,5){\makebox(0,0)[lc]{$q_n$}}
\put(-10,5){\makebox(0,0)[rc]{$q_1$}}

\put(25,-20){\oval(10,30)[]}
\put(-25,-20){\oval(10,30)[]}
\linethickness{2.pt}
\put(30,-20){\line(1,0){20}}
\put(-30,-20){\line(-1,0){20}}
\linethickness{0.4pt}

\put(-20,-10){\line(1,0){40}}
\put(-20,-20){\line(1,0){40}}
\put(-20,-30){\line(1,0){40}}

\end{picture}
\ec

\vspace{1cm}

\bc
Fig.~3
\ec


\begin{thebibliography}{99}

\bibitem{1} M.Braun and V.Vechernin,
{\it Nucl. Phys.} {\bf B427} (1994) 614.

\bibitem{2} W.P.Sh\"{u}etz {\it et al.,
Phys. Rev. Lett.}  {\bf 38} (1977) 259.

\bibitem{3} A.M.Baldin {\it et al.,
Preprint P1-11168}, (JINR, Dubna, 1977);\\
A.M.Baldin {\it et al.,
Preprint E1-82-472}, (JINR, Dubna, 1982).

\bibitem{4} N.A.Nikiforov {\it et al.,
Phys. Rev.} {\bf C22} (1980) 700.

\bibitem{5} BCDMS collaboration,
{\it Z. Phys.} {\bf C63} (1994) 29.

\bibitem{6} A.V.Efremov, A.B.Kaidalov, G.I.Lykasov and N.V.Slavin,
{\it Yad. Fiz.} {\bf 57} (1994) 932.

\bibitem{7} A.B.Kaidalov, {\it Phys. Lett.} {\bf B116} (1982) 459;\\
A.B.Kaidalov and K.A.Ter-Martirosyan, {\it Phys. Lett.} {\bf 117} (1982) 247.

\bibitem{8} M.A.Braun, {\it Nucl. Phys.} {\bf A523} (1991) 694.

\bibitem{9} M.S.Dubovikov and K.A.Ter-Martirosyan,
{\it Nucl. Phys.} {\bf B124} (1977) 163;\\
A.Kaidalov, L.Ponomarev and K.Ter-Martirosyan,
{\it Yad. Fiz.} {\bf 44} (1986) 722.

\bibitem{10} V.N.Gribov,
{\it Zh. Eksp. Teor. Fiz.}, {\bf 56} (1969) 892.

\bibitem{11} A.B.Kaidalov and K.A.Ter-Martirosyan,
{\it Nucl. Phys.} {\bf B75} (1974) 471.

\bibitem{12} S.J.Brodsky and H.J.Lu,
{\it Phys. Rev. Lett.} {\bf 64} (1990) 1342.

\bibitem{13} N.A.Amos {\it et al}, {\it Nucl. Phys.} {\bf B262} (1985) 689.

\end{thebibliography}
\end{document}